# Ferromagnetic Josephson switching device with high characteristic voltage

Timofei I. Larkin,[1] Vitaly V. Bol'ginov,[1] Vasily S. Stolyarov,[1] Valery V. Ryazanov,[1,a)] Igor V. Vernik,[2,b)] Sergey K. Tolpygo,[2] and Oleg A. Mukhanov[2]

[1] *Institute of Solid State Physics, Russian Academy of Sciences, Chernogolovka, 142432, Russia*

[2] *HYPRES Inc., 175 Clearbrook Road, Elmsford, New York 10523, USA*

We develop a fast Magnetic Josephson Junction (MJJ) – a superconducting ferromagnetic device for a scalable high-density cryogenic memory compatible in speed and fabrication with energy-efficient Single Flux Quantum (SFQ) circuits. We present experimental results for Superconductor-Insulator-Ferromagnet-Superconductor (SIFS) MJJs with high characteristic voltage $I_cR_n$ of >700 μV proving their applicability for superconducting circuits. By applying magnetic field pulses, the device can be switched between MJJ logic states. The MJJ $I_cR_n$ product is only ~30% lower than that of conventional junction co-produced in the same process, allowing for integration of MJJ-based and SIS-based ultra-fast digital SFQ circuits operating at tens of gigahertz.

---

a) Also at InQubit Inc., 21143 Hawthorne Blvd., Torrance, CA 90503, USA
b) Author to whom correspondence should be addressed. Electronic mail: vernik@hypres.com

High speed, low power superconducting Rapid Single Flux Quantum (RSFQ) digital circuits have already found their applications in Digital-RF systems impacting communications and signal intelligence applications [1, 2]. Recently, a new energy-efficient generation of RSFQ circuits, eSFQ and ERSFQ logics, offered a way to overcome the low energy efficiency of conventional technologies for the next generation of supercomputers [3]. However, the practical applications of these superconducting digital technologies will inevitably be very limited without compatible in speed and signal levels, high-capacity, energy-efficient Random Access Memory (RAM). The largest superconducting RAM demonstrated to date, a 4 kbit RAM [4], is insufficient for practical applications and hardly compatible with SFQ-type circuits.

The low density of superconducting memory is directly related to a relatively large size of memory cells based on SFQ storage loops coupled to address lines via transformers which are difficult to scale [5-8]. The required *ac* power posed an additional implementation problem for achieving larger capacity RAM integrated circuits [8]. In order to get around the low capacity of superconductor RAMs, hybrid superconductor-semiconductor schemes were pursued [9, 10]. However, these approaches can address only limited applications and cannot satisfy the need for a fast, energy efficient memory in a close proximity to the digital circuits, preferably on the same chip. Alternatively, combining superconducting elements with ferromagnetic layers and dots was suggested to achieve higher density of superconducting memory [11, 12]. However, these ideas did not go beyond initial concepts nor address compatibility with SFQ circuits.

Recently, we have introduced a memory cell based on Magnetic Josephson Junction (MJJ), a Josephson switching device with ferromagnetic (F) layer(s). The MJJ critical current can change and retain its value by ferromagnet magnetization, so that a memory element size is defined by the scalable small MJJ device [13]. With achieving MJJ switching speed comparable to that of conventional JJs, both types of junctions can be integrated into a single circuit operating in an SFQ non-hysteretic switching regime, enabling a low power and high speed memory operation. Since such memory will be electrically and physically compatible with SFQ-type circuits, it enables a fabrication of memory and digital circuits in a single process on the same chip including a high density 3D integration.

In order to achieve a high energy-efficiency of MJJ-based memory comparable to that of digital SFQ-type circuits, the MJJs must comply with two main requirements: fast and low-energy F-layer remagnetization for *Write* operation and fast SFQ junction switching for *Read* operation. In this paper, we focus on the development of the MJJ with high characteristic voltage - the key device enabling fast non-destructive readout and low-power write for an energy-efficient and high-capacity, non-volatile, and high-speed memory compatible to SFQ circuits.

It was already proven, that a switching MJJ can be formed as a Superconductor-Ferromagnetic-Superconductor (SFS) junction. For the ferromagnetic layer, a weak and magnetically soft PdFe alloy with low Fe-content was chosen [14]. The ability of Nb/Pd$_{0.99}$Fe$_{0.01}$/Nb SFS junction to operate as a switch was demonstrated in Institute of Solid State Physics (ISSP) in 2010 [15]. Specifically, an application of small external magnetic field changed the magnetization of the ferromagnetic layer that in turn changes the junction $I_c$, allowing the realization of two distinct states with high and low $I_c$, corresponding to logical

"0" and "1" states, respectively. However, the characteristic voltage $I_cR_n$ of these SFS devices was in the order of nanovolts, which makes them too slow (~MHz rate) to be applicable to prospective memory applications.

The MJJs have to be comparable in switching speed with conventional SIS JJs. By inserting an additional isolation tunnel layer $I$ in the junction (i.e., fabricating an SIFS structure), one should be able to increase $V_c = I_cR_n$ to ~1 mV to achieve high switching frequency. This should bring the MJJ switching speed close to that of SIS JJs, while retaining useful memory properties of SFS MJJs.

Initial attempts to implement SIFS tunnel junctions were done with a hard-magnetic barrier (Ni) [16]. However, the remagnetization of such F-layer required too much energy. Another ferromagnet, CuNi alloy, was used as a ferromagnetic layer in superconducting SFS phase shifters in both the "classical" digital and quantum circuits. A Toggle Flip-Flop (TFF) with the embedded SFS π-JJ [19] as a non-switching phase shifting element inserted into a storage loop [17] and a π-biased phase qubit [18] were demonstrated. The CuNi alloy is highly suitable for such constant phase shifters due to its stable magnetic domain structure (high coercive field) and out-of-plane magnetic anisotropy. However, a ferromagnetic layer with an in-plane magnetic anisotropy and small coercive field would be more convenient for the implementation of switching devices. A magnetically soft PdFe alloy with low Fe-content matches these requirements. The SIFS MJJ devices based on Nb-Al/AlO$_x$-Pd$_{0.99}$Fe$_{0.01}$-Nb tunnel junctions with $V_c$ from 100 to 400 μV were reported in [13]. In this paper, we present further development and investigation of these devices.

Our fabrication process is based on a co-fabrication approach, which was used in the past for fabrication RSFQ circuits using different lithography capabilities [20]. This time, we undertake a more challenging task of co-fabricating active devices using a combination of HYPRES and ISSP fabrication processes. HYPRES standard 4.5 kA/cm$^2$ SIS process [21] is used to fabricate high speed SFQ digital and mixed-signal integrated circuits (ICs). Therefore, the ability to fabricate MJJ devices based on this process proves the material and physical compatibility between digital and perspective memory fabrication processes. Because of an extensive experience of the ISSP group with SFS junctions, the fabrication was split into two major steps. Firstly, we produced a series of 150-mm wafers with an in-situ deposited Nb-Al/AlO$_x$-Nb trilayers with 4.5 kA/cm$^2$ target Josephson critical current density. The wafers were then diced into 15x15 mm$^2$ samples and transferred to the ISSP for subsequent deposition of a ferromagnetic layer (Pd$_{0.99}$Fe$_{0.01}$) and top Nb counter electrode. The resultant structure is of SI(S)FS type, in which superconductivity of (S)-layer is expected to be substantially suppressed by the adjacent ferromagnet and, therefore, be close to the desired SIFS configuration. For the SI(S) fabrications, two thicknesses of the counter electrode were used, 20 nm and 15 nm whereas Nb base electrode thickness was 120 nm for all samples studied. At the ISSP facility, the samples were cleaned in acetone/methanol/IPA and blow dried with N$_2$ gas. In-situ Ar sputter etching was used to remove about 10 nm of Nb oxide layer before depositing ferromagnetic layer. The PdFe/Nb bilayer was deposited using rf- and dc-magnetron sputtering. The PdFe layer was thin enough (14-18 nm) to avoid significant critical current suppression. The top Nb layer thickness was about 150 nm to ensure uniform supercurrent flow through a Josephson junction. Then, we formed a square mesa of 10x10 μm$^2$ sizes by photolithography, reactive ion etching (RIE) of the top Nb layer and argon plasma etching of PdFe and Al/AlOx layers, and patterned the bottom Nb-electrode with photolithography and RIE. At the third step, we formed an isolation layer with a contact (wiring) window by using thermal evaporation of SiO and a lift-off process. Junction contact size in SiO layer was 4x4 μm$^2$. At the last step we formed Nb wiring electrode of 450 nm thickness using magnetron sputtering and a lift-off. We used argon RF-etching to ensure a good interface transparency between the wiring and the top Nb electrode of the mesa. For a reference, we also produced conventional SI(S)S junctions on the samples from the same wafers by excluding the F-layer deposition step.

It is difficult to asses if superconductivity in the (S)-layer of the SI(S)FS junctions is totally suppressed by a proximity effect or it is partially remains in a superconducting state. In the latter case, the resultant structure would be close to a stacked combination of SIS and SFS junctions. In this paper, we did not focus on a systematic investigation and comparison of all possible combinations. We believe that for the purpose of the memory applications discussed above, both structures might be applicable. As we will further describe below, we were able to observe a non-volatile magnetization and corresponding change of junction critical current as desired.

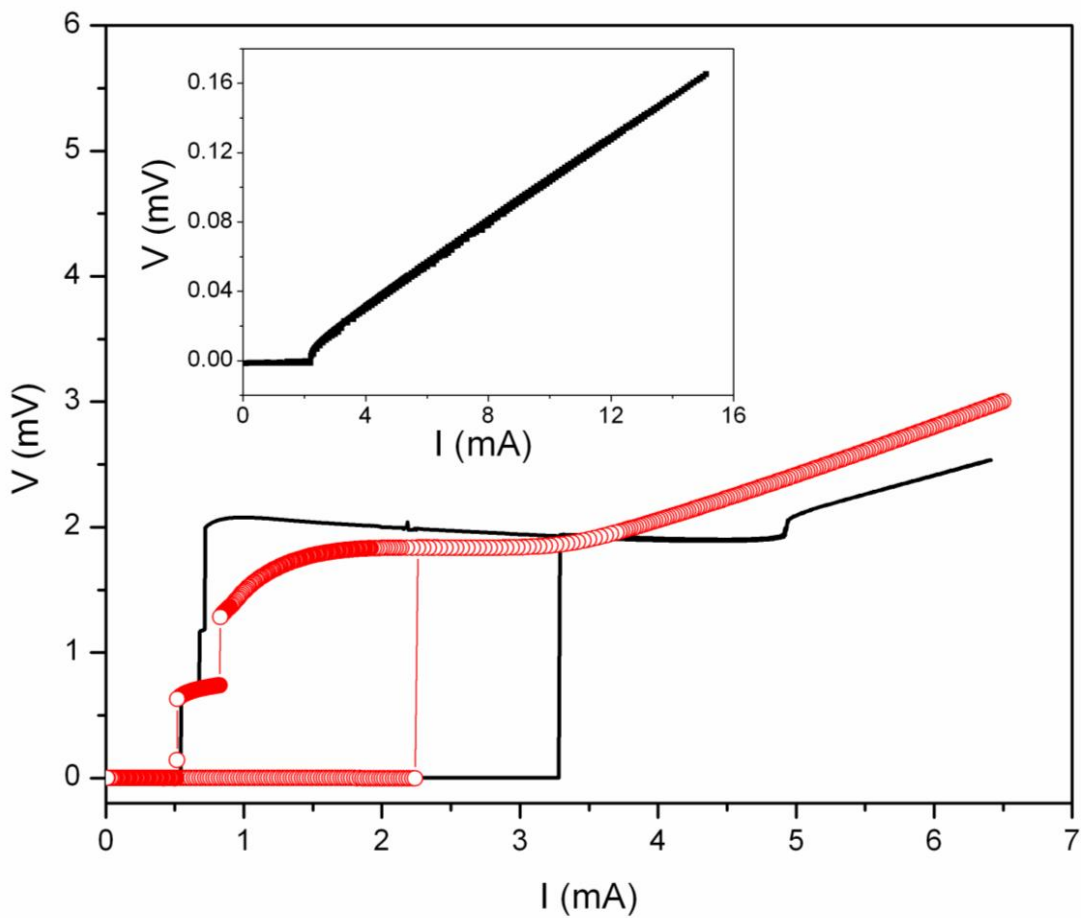

*Fig. 1. Current-voltage characteristics of SI(S)FS (open circles) and reference SI(S)S (line) with 14-nm ferromagnetic layer Josephson junctions fabricated on different samples from the same wafer. Inset shows a current-voltage characteristic of a SI(S)FS Josephson junction externally shunted with a small resistor.*

All our measurements were performed in a variable temperature liquid He cryostat at 4.2 K. The sample holder was placed in a vacuum can with He gas added for better heat exchange. The sample temperature was controlled using a carbon thermometer and a compact heater wounded by a twisted pair and glued close to the sample. Figure 1 shows the measured current-voltage (I-V) characteristics of SI(S)FS and SI(S)S junctions produced in the similar fabrication cycles. Only a positive quadrant is shown for simplicity. The I-V curve of the reference SI(S)S junction shows critical current $I_c$ = 3.3 mA which is very close to 70% of a current step at the gap voltage, indicating a uniform bias current distribution. This I-V-curve exhibits somewhat suppressed superconducting gap of ~2 mV that is less than 2.6 mV gap voltage observed for JJs made by the standard HYPRES process. We attribute this to a gap suppression in the thin Nb counter electrode (S) which is 150-200 nm thinner here than that in the standard HYPRES process, due to its possible contamination or damage during post trilayer deposition steps (e.g., wafer dicing into chips, Ar sputter etching to remove Nb oxide, etc.). A step at ~ 700 µV observed for both SIS and SIFS I-V curves can be attributed to the well-known peculiarity at the gap difference voltage due to different quality of the base and counter Nb electrodes.

The observed $I_c$ corresponds to critical current density $J_c$ of ~3.3 kA/cm$^2$, consistent with the target $J_c$ of 4.5 kA/cm$^2$ and the reduced gap in Nb counter electrode. The negative slope of the I-V curve visible in Fig. 1 may be explained by a non-equilibrium state of SIS junctions observed earlier [22, 23]. This is consistent with the uniform change in the energy gap under the influence of quasiparticle injection [24]. It is also possible to attribute this to junction self-heating due to large size of JJs, since this type of I-V curves has never been observed in JJs with small sizes (less than ~16 µm$^2$) produced by the standard HYPRES process and the similar critical current density.

Both the gap voltage and $I_c$ are suppressed for the SI(S)FS junction fabricated with a 14 nm PdFe layer. The $I_c R_n$ product of SI(S)FS junction is suppressed by ~30% with respect to SI(S)S. This makes the integration of such junctions straightforward in superconducting SFQ digital circuits by proportionally scaling their geometrical dimensions. Inset to Fig. 1 shows an I-V curve of the same SI(S)FS junction that was shunted externally by a small ~ 0.01 Ω resistor made of Al wire. This shunting was performed post-process to obtained a device with nonhysteretic I-V curve and simplify automated $I_c(H)$ measurements. The inset I-V curve resembles the typical I-V for shunted SIS Josephson junctions used in RSFQ circuits, although with much smaller $I_c R_{shunt}$ product. To use a full potential of $I_c R_{shunt}$ product offered by SI(S)FS, the deposition of shunt resistors needs to be further included in junction fabrication process like it is done in the standard RSFQ fabrication [21].

Figure 2 demonstrates a typical Fraunhofer-like $I_c(H)$ dependence of MJJ critical current on external magnetic field $H_{ext}$ aligned parallel to the layers within the MJJ and supplied by an external solenoid. For each measured point of the $I_c(H)$ curve, the current through the sample is swept from a subcritical value upwards until the threshold voltage is exceeded while the external magnetic field is fixed. This measurement was performed for the shunted MJJs described above, since they do not require the reduction of the bias current to almost zero for each $I_c(H)$ measurement point. In addition, the shunted MJJ is much less susceptible to the external noise. The observed offset of ~0.5 mA in $I_c(H)$ curve (Fig. 2) is related to our measurement technique, in which $I_c$ is defined as the current at which threshold voltage of ~5 μV across the MJJ with $R_{shunt} \sim 0.01\Omega$ is reached (see inset to Fig. 1). For a comparison, we did not observe any offsets of $I_c(H)$ for the unshunted MJJs.

The $I_c(H)$ curve for a shunted MJJ has a noticeable hysteresis (Fig. 2). The starting point of our measurements is from the state with an applied external magnetic field of ~10 Oe, and not from a demagnetized state. The external field is swept from 10 Oe to -17 Oe and back, and the corresponding magnetic flux through the junction, $\Phi$ follows solid and dashed arrows, respectively. The $I_c(H)$ dependence with $H$ decreasing is shown by solid circles and solid line and $I_c(H)$ with $H$ increasing is shown by open circles and dashed line. For the measured MJJs, a self-field effect is non-negligible. When a bias current is applied to the MJJ, it produces $H_{curr}$ that significantly influences $I_c$ and adds to the applied external field. This self-field effect is exhibited in inset (a) to Fig. 2 showing a symmetric with respect to the origin dual-polarity $I_c(H)$ curve with main $I_c$ maximum at non-zero $H_{ext}$. In this case the effective magnetic field seeing by the ferromagnetic layer inside of the MJJ is $H_{eff} = H_{ext}+H_{curr}$. Coupled with a hysteretic nature of $M(H)$, this results in a hysteretic nature of $\Phi$ vs. $H$ dependencies (where $M$ is magnetization of the F-layer) as shown in insets (b) and (c) to Fig. 2. Inset (b) shows a regular hysteresis loop as a function of external field with zero bias current and therefore zero $H_{curr}$. Here, the maximum value of $I_c$ is reached at points 1 and 2, where the total magnetic flux through the junction equals zero. Inset (c) demonstrates a transformation of the loop of inset (b) after current of 2 mA (close to $I_c$) is applied to the junction. Point 4 in inset (b) moves to point 4' and now corresponds to zero flux through the junction, and thus a maximum possible critical current, while the loop itself shifts horizontally. This creates the desired offset to make two $I_c(H)$-branches asymmetric relative to the y-axis.

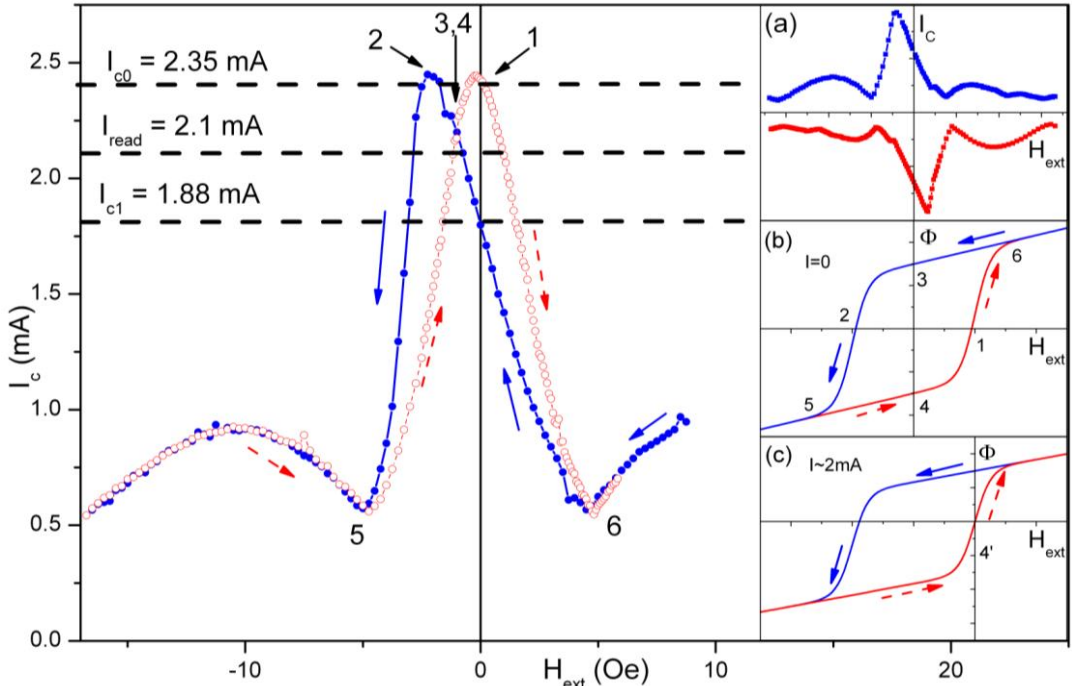

*Fig. 2. Hysteretic dependence of critical current $I_c$ on magnetic H field for Nb-Al/AlOx-Nb-PdFe-Nb shunted SI(S)FS MJJ with a 14 nm ferromagnet layer. Magnetic field sweep directions are shown by arrows. Insets show dual-polarity Ic(H) and magnetization curves $\Phi(H)$.*

From Fig. 2 one can clearly see that the SI(S)FS junction critical current, $I_c$, depends on magnetic prehistory. The application of an external magnetic field changes the magnetization of the ferromagnetic layer that in turn changes the junction $I_c$ allowing the realization of two distinct states with high and low $I_c$, corresponding to logical "0" and "1" states, respectively, i.e., states with $I_{c0}$ = 1.88 mA and $I_{c1}$ = 2.35 mA in Fig. 2. Thus, one can choose a junction bias current ($I_{read}$ = 2.1 mA in Fig. 2) to switch the SI(S)FS junction from a superconducting to a resistive state by a pulse of magnetic field. This experiment is presented in Fig. 3, where positive and negative magnetic field pulses switch the SI(S)FS junction from a superconducting (zero-resistance) to a resistive state and back. Note, that our data acquisition program turns off the read current during the application of magnetic pulses, appearing as "switching to zero" visible at ~160 - 170 sec. When the junction is in a superconducting state, one cannot distinguish such $I_{read}$ turning off. In order to investigate a magnetization stability of the MJJ, we applied repetitive magnetic pulses of the same polarity. One can see that the repetition of magnetization pulses does not further magnetize the MJJ and depress its $I_c$. The behaviour of MJJ is well defined: a positive pulse switches it to state "1", and a negative pulse switches it to "0" state, regardless of prior MJJ state.

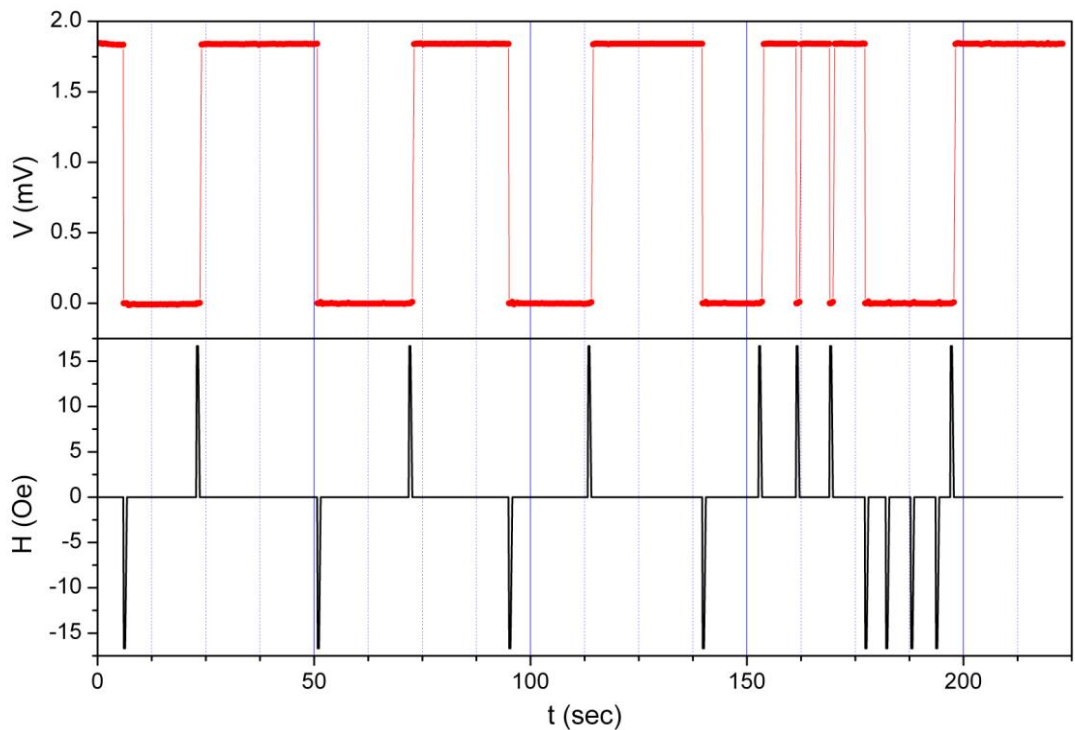

*Fig. 3. Switching of SI(S)FS MJJ between "0" and "1" states by remagnetization with external magnetic field. V(t) – average junction voltage, H(t) - applied magnetic field.*

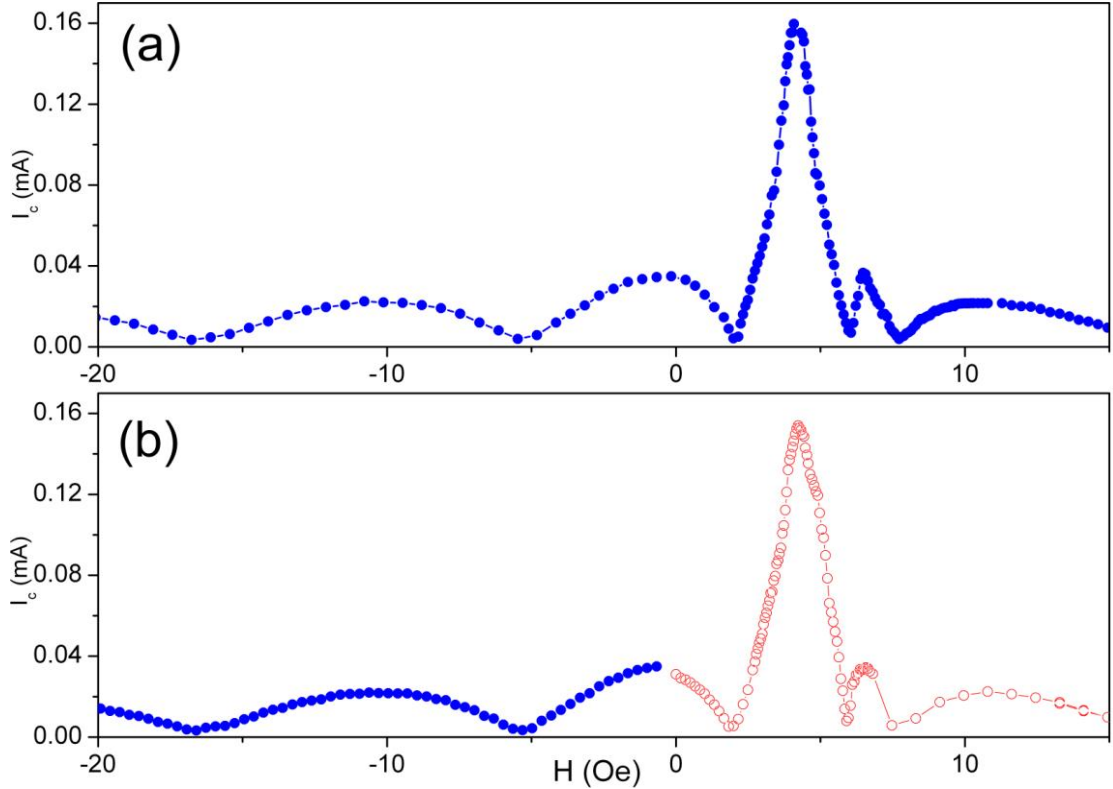

*Fig. 4. The $I_c(H)$ dependencies for a SI(S)FS MJJ verifying a memory retention over 7 hours. (a) $I_c(H)$ curve with the external magnetic field swept from -20 Oe to 15 Oe. (b) $I_c(H)$ curve with the external magnetic field increased from -20 Oe to 0 Oe (closed circles) and then after a 7 hour hold time further increased from 0 Oe to 15 Oe (open circles).*

Figure 4 addresses the retention ability of SI(S)FS MJJ as a memory element. These $I_c(H)$ measurements were performed for an MJJ with ferromagnetic PdFe layer with thickness $d_F$ = 18 nm, 4 nm thicker then for the MJJ used in Figs. 1-3. Expectedly, this MJJ exhibits lower $I_c$ =158 µA at 4.2 K. The $I_c(H)$ dependencies in Fig. 4 consist of three data sets. First, Fig. 4a shows $I_c$ measurements when the external magnetic field was swept from -20 Oe to 15 Oe. The second data set is shown in Fig. 4b (by solid circles) with the external magnetic field changed from -20 Oe to 0 Oe. Then the magnetized MJJ was left for 7 hours at 4.2 K, well below Curie temperature estimated for this sample to be around 10 K. Finally, the third set was obtained after this 7 hour hold with the magnetic field further increased from 0 Oe to 15 Oe (see Fig. 4b, open circles). The $I_c(H)$ curve in Fig. 4b convincingly shows the continuity between the second and the third data sets, proving that SI(S)FS MJJ retains its magnetization for at least 7 hours.

In summary, a perspective switching devices with high characteristic speed for high capacity, energy-efficient cryogenic memory compatible with energy-efficient SFQ-type logic have been fabricated and measured. Using a co-fabrication approach at HYPRES and ISSP, we demonstrated SI(S)FS (Nb-Al/AlO$_x$-Nb-PdFe–Nb) junctions with high characteristic voltage $I_cR_n$ of ~700 µV, which is just ~30% lower of that of the co-fabricated SI(S)S (Nb-Al/AlO$_x$-Nb-Nb) junctions. This makes these junctions the fastest MJJs demonstrated to date and fairly compatible to the speed of conventional SIS JJs used in RSFQ circuits. We have experimentally proved that critical current of these MJJs can change and retain its value by the ferromagnet layer magnetization, so that a memory cell size is now defined by a scalable MJJ device, which leads to a high-density RAM. Our experiments showed a non-volatile and non-destructive readout of the MJJ memory element. By fabricating MJJs starting with standard HYPRES process used for digital SFQ circuits, we proved the fabrication compatibility for a future integration of digital and memory circuits on a single chip.

We thank R. Hunt, J. Vivalda, V. Shilov, N. Stepakov, V. Oboznov, M. Paramonov and V. Shkolnikov for assistance in sample preparation and measurements, M. Kupriyanov, A. Kadin, I. Nevirkovets for useful discussions, and M. Manheimer and S. Holmes for encouragement.